\documentclass[sigconf, preprint, nonacm]{acmart}

\usepackage{graphicx}
\usepackage{subfig}
\graphicspath{ {./Figures/} }

\begin{document}

\title{The Importance of Low Latency to Order Book Imbalance Trading Strategies}

\author{David Byrd}
\email{db@gatech.edu}
\affiliation{
    \institution{School of Interactive Computing \\
                 Georgia Institute of Technology}
    \city{Atlanta}
    \state{Georgia}
    \postcode{30308}
}
\author{Sruthi Palaparthi}
\email{sruthi.p@uga.edu}
\affiliation{
    \institution{Department of Computer Science \\
                 University of Georgia}
    \city{Athens}
    \state{Georgia}
    \postcode{30303}
}
\author{Maria Hybinette}
\email{maria@cs.uga.edu}
\affiliation{
    \institution{Department of Computer Science \\
                 University of Georgia}
    \city{Athens}
    \state{Georgia}
    \postcode{30303}
}
\author{Tucker Hybinette Balch}
\email{tucker@cc.gatech.edu}
\affiliation{
    \institution{School of Interactive Computing \\
                 Georgia Institute of Technology}
    \city{Atlanta}
    \state{Georgia}
    \postcode{30308}
}

\begin{abstract}
There is a pervasive assumption that low latency access to an exchange is a key factor in the profitability of many high-frequency trading strategies. This belief is evidenced by the ``arms race'' undertaken by certain financial firms to co-locate with exchange servers.
To the best of our knowledge, our study is the first to validate and quantify this assumption in a continuous double auction market with a single exchange similar to the New York Stock Exchange.
It is not feasible to conduct this exploration with historical data in which trader identity and location are not reported.
Accordingly, we investigate the relationship between latency of access to order book information and profitability of trading strategies exploiting that information with an agent-based interactive discrete event simulation in which thousands of agents pursue archetypal trading strategies.
We introduce experimental traders pursuing a low-latency order book imbalance (OBI) strategy in a controlled manner across thousands of simulated trading days, and analyze OBI trader profit while varying distance (latency) from the exchange.
Our experiments support that latency is inversely related to profit for the OBI traders, but more interestingly show that latency \emph{rank}, rather than absolute magnitude, is the key factor in allocating returns among agents pursuing a similar strategy.
\end{abstract}


\begin{teaserfigure}
  \includegraphics[trim=19cm 7cm 30cm 6cm, clip, height=1in, width=\textwidth] {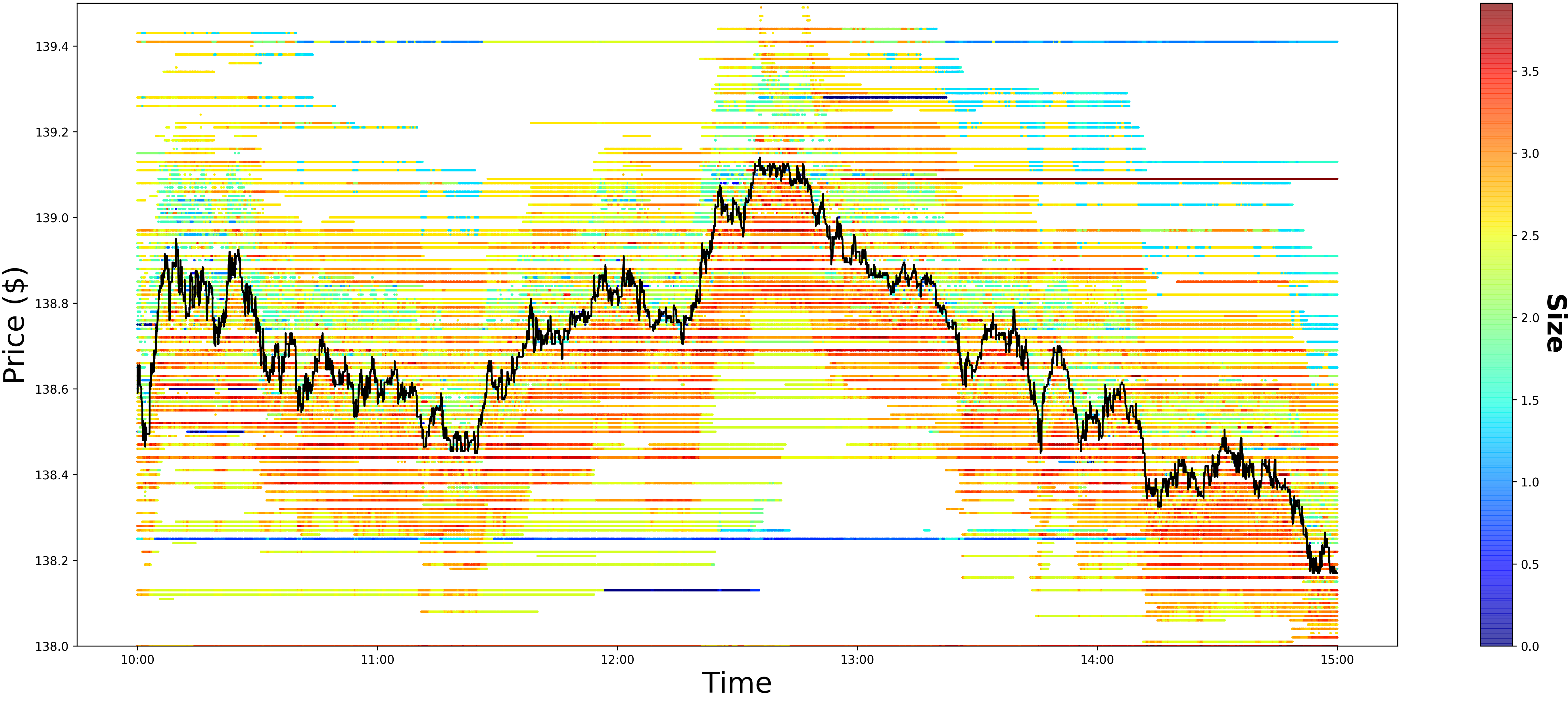}
\end{teaserfigure}

\maketitle

\section{Introduction}

It is commonly understood that execution speed is essential for a successful high-frequency trading (HFT) strategy.  In \emph{High Frequency Trading: A Practical Guide}, Irene Aldridge observes that ``High-frequency trading relies on fast, almost instantaneous, execution of orders'' and that ``even a second’s worth of delay induced by hesitation or distraction on the part of a human trader can substantially reduce the system’s profitability'' \citep{aldridge2013high}.  Although there is a well-developed body of financial literature around such strategies, works that rely on analysis of historical market data cannot reliably quantify the precise, continuous benefit of specific latency levels to a strategy.  Publicly available market information does not identify individual actors or provide a mechanism to reveal the latency of an actor's order execution after a trading decision has been reached, so it is not feasible to use such data to infer the importance of latency for such strategies.

To address this problem, we employ an agent-based interactive discrete event simulation to construct a market with nanosecond precision to fully account for agent computational delay and communication latency.  We experiment within that simulated market to assess the effect of low latency on overall agent returns for a typical high-frequency strategy.  As far as we know, this is the first published study to directly and quantitatively assess the impact of variable latency on a trader's profit.

Our work specifically focuses on market microstructure, or making short-term directional trading decisions based on observed quotes.  Richard Lyons divides this group of High Frequency Trading (HFT) strategies in two sets \cite{lyons2001microstructure}, both relating to mismatches in supply and demand:

\begin{itemize}
    \item \textbf{Inventory effects} pertaining to a market maker absorbing inventory and attempting to dissipate risk.
    \item \textbf{Portfolio balance effects} in which the order flow produces a more lasting impact.
\end{itemize}

We introduce a group of experimental agents following a strategy from the first category, inventory effects, and investigate the effect of absolute and relative latency on the profitability of each such agent.

\section{Background and Related Work}

Our work draws from prior lines of financial research in market microstructure trading and latency arbitrage, and computational research in discrete event simulation and agent-based modeling.  Here we describe some of the important prior work in each area.

\subsection{Market Microstructure Trading}

Lawrence Harris described three types of stylized traders relevant to market microstructure trading: \cite{harris1998optimal}

\begin{enumerate}
    \item \textbf{Liquidity traders} are uninformed traders whose execution timelines are externally motivated by a client demand or a need to modify cash holdings.
    \item \textbf{Informed traders} are those who have specific, private information they believe correlates with short term price movement, which must be acted on quickly.
    \item \textbf{Value-motivated traders} have an exogenous opinion on a stock's true worth and are motivated to buy or sell at specific prices that represent significant deviations from that value.
\end{enumerate}

Harris further identifies informed traders as \emph{aggressive}, utilizing market orders or limit orders near the spread to ensure their private information is monetized before its expiration.  Value-motivated traders are characterized as more \emph{passive}, placing limit orders far from the spread to execute only if the order flow reaches their notion of a fair price.

Our study includes simulated versions of these  three stylized strategies.  We construct an environment of informed traders and value-motivated traders that serve as ``background'' market agents then, under various conditions, evaluate the performance of a particular kind of liquidity trader which attempts to predict short-term price changes using an order book imbalance (OBI) indicator.

Bloomfield et al constructed experimental markets to test some of Harris's predictions, finding that informed traders take liquidity when the gap between current prices and those suggested by their time-sensitive information is high, but provide liquidity when that gap is low \cite{bloomfield2005make}.  This provides empirical support for our simulated liquidity traders' belief that a large amount of liquidity provision near the spread indicates impending directional movement.

\subsection{Latency Arbitrage}

Wah and Wellman have previously studied latency arbitrage with a simulated two-market model plus public price quotes from an NBBO (National Best Bid and Offer) provider \cite{wah2013latency}.  They found that the presence of a high frequency trading (HFT) agent arbitraging the two markets negatively impacted the surplus achieved by other traders in excess of the amount it obtained for itself.  The latency arbitrageur did improve order execution speed (a common defense of HFT activity) but actually caused a \emph{wider} bid-ask spread.  

Our work examines a different aspect of latency arbitrage, introducing multiple competing liquidity traders which pursue a low-latency strategy with a single exchange.  We focus on the relationship between absolute and relative communication latency levels and the profitability of each liquidity trader.

\subsection{Discrete Event Simulation}

We conduct our experiments in an event driven framework built on a discrete event simulation (DES) system kernel.  Under a DES model, the system changes state only at the edges of discrete points in time.  Conventional approaches to DES are either time driven (synchronous) or event driven (asynchronous).  

In time driven (or time ``stepped'') systems, progress is driven by incrementally advancing time, which is typically represented by a global counter.  The counter is increased by a fixed amount, the minimum resolution, and then events that have a time stamp matching the current counter are processed.  One disadvantage of time stepped simulation is the wasting of computation time on steps during which no activity takes place.  For example, when exploring high frequency trading (HFT), it might be important to allow agents to act with nanosecond time resolution.  If the counter is at time $t$ after some agent places an order, and the next agent will act just fifty microseconds later, the system will have to sequentially check for activity at time steps $t+1$ to $t+49999$ even though there will be no state changes.

Event driven systems mitigate the wasted computation time caused by such ``idling'' by changing the mechanism of time progression.  Instead of incrementing a counter, the system's simulated time jumps to the time stamp of the next (chronologically earliest) scheduled event.  Progress through time is guaranteed under the assumption that until the simulation ends, there will always be \emph{some} next scheduled event.  Under this model, each time an agent acts, it will schedule its next action through a priority queue in the system kernel.

To revisit the above HFT example in the event driven context, when an agent places an order at time $t$ and the next agent will act fifty microseconds later, the kernel's event priority queue will have time stamp $t+50000$ as the first entry, and so the system time can transition directly from $t$ to $t+50000$ with no consideration of the time between.  This can be done because the system state changes only through agent actions, and no action is scheduled during the skipped interval.  Event driven simulation is efficient precisely because of this feature, that all time in between state changes is safely ignored.  Event driven kernels are also amenable to parallelization of computation that further speeds execution, as demonstrated in Optimistic Simulation Time Warp \cite{jefferson1982fast} and in Conservative Simulation \cite{chandy81}.  DES has been widely adopted and used to address important problems in health care \cite{karnon12}, supply chain management \cite{tako2012application}, manufacturing \cite{negahban2014simulation}, and financial markets \cite{jacobs04,jacobs10,levy94,wah2013latency}.

Both Levy et al \cite{levy94} and Wellman and Wah \cite{wah2013latency} take a synchronous approach to simulation, but for our larger scale experiments we believe that an asynchronous simulation provides greater flexibility and scalability.  This view is shared by Jacobs et al \cite{jacobs04,jacobs10}, who introduced a financial simulation framework called JLMSim.  JLMSim is a discrete event simulator that incorporates trading rules for simple strategies and reproduces the changes in the market by executing buy and sell orders from an order book, but with the limitation that it does not support the implementation of complex custom trading strategies.

\subsection{Agent Based Modeling}

Our system uses an agent based model (ABM), formed by a set of autonomous agents that interact with their environment, including other agents, to achieve their objectives.  Agent based modeling and simulation (ABMS) has been successfully employed in a variety of application domains \cite{macal2009agent} such as social science \cite{axelrod97}, computational economics \cite{tesfatsion2006handbook}, and marketing \cite{negahban2014simulation}.  We follow prior work in the area by modeling a stock market based on the behavior of individual investors \cite{levy94} represented as strategic trading agents.  Agent-based financial markets have been shown to be effective for dynamic situations in which investor behavior must change in response to the environment \cite{lebaron2011active}.

Some notable agent-based simulators from the literature are Swarm \cite{minar1996swarm}, MASON \cite{luke2005mason}, and Player/Stage \cite{gerkey2003player}.  There are also simulators like SASSY \cite{Hybinette06} that support large numbers of agents through an optimistic parallel kernel.  While our system is not a parallel simulation kernel we draw inspiration from their design and currently support many thousands of agents.

\section{Approach}

To test the hypothesis that lower communication latency with an exchange will correlate with higher trading period profits for order book imbalance (OBI) liquidity traders, we construct an agent-based interactive discrete event simulation using components described in Background and Related Work.  The simulation provides a Kernel which enforces the proper flow of time and through which all inter-agent communication must occur, and the simulation environment represents a modern electronic stock market in which numerous strategic trading agents place bids and offers with a single exchange agent.

\begin{figure}[!h]
    \includegraphics[width=3in]{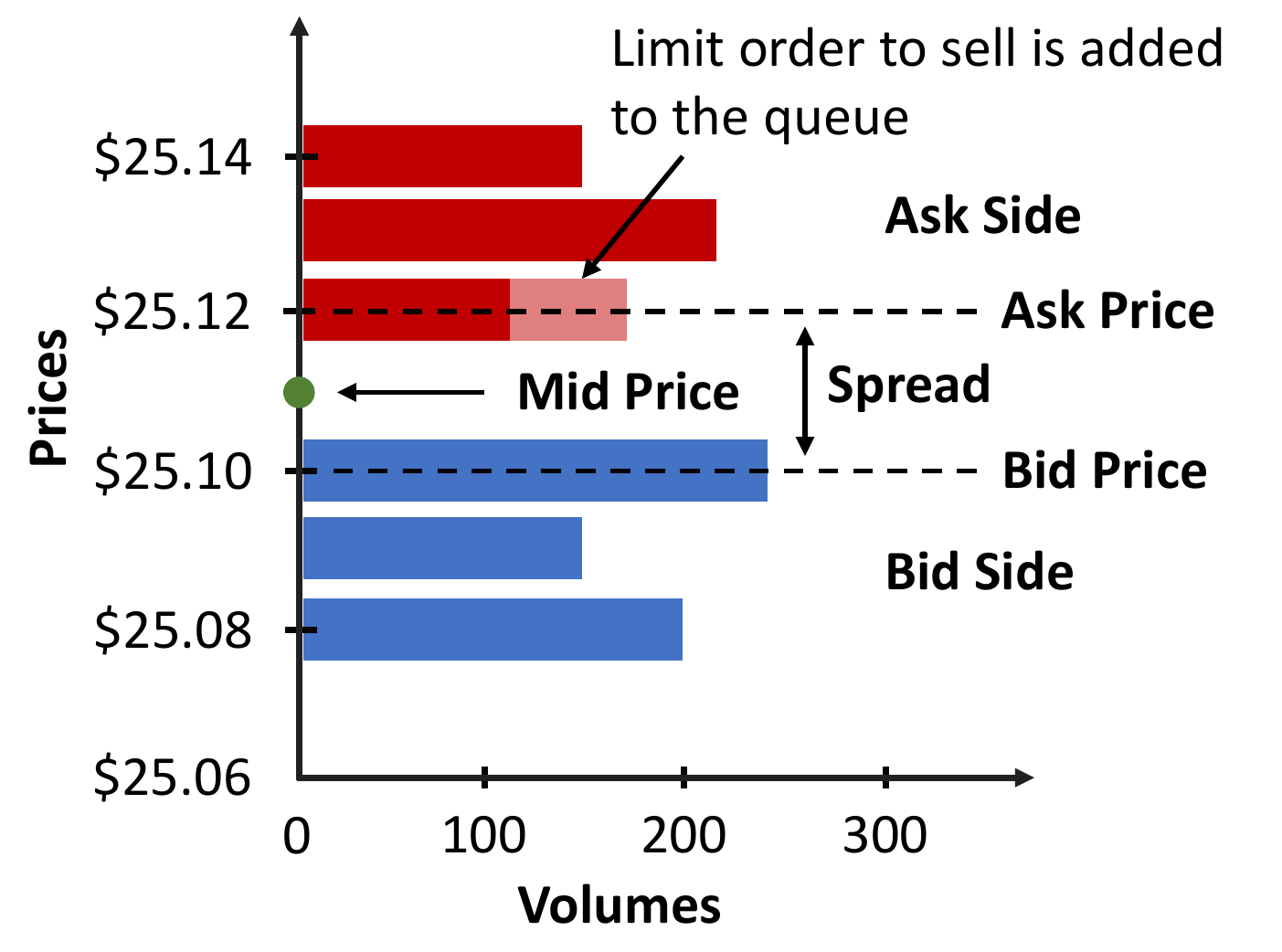}
    \caption{Example illustrating liquidity provisioned to a limit order book.}
    \label{fig:lob_chart}
\end{figure}

\subsection{Market Simulation}

At the core of our simulated market is an exchange agent which accepts orders to buy (bid prices) and sell (ask prices) specified quantities of various securities.  Such orders may optionally contain an additional \emph{limit price} which prohibits transaction at any less advantageous price for the submitting agent.  Orders with a limit price, called \emph{limit orders}, may not immediately transact, and will instead be recorded into the \emph{limit order book} for the relevant security as shown in Figure \ref{fig:lob_chart}, to await future transaction if a matching order should arrive.  Orders without a limit price, called \emph{market orders}, have no such restriction and will always transact immediately at the best currently available price.  

As the limit order book consists of all unfulfilled orders to transact a security, the limit prices and quantities of visible bid and ask orders can be interpreted as some representation of the supply and demand for the security at one moment in time.  Measurable aspects of the limit order book include: the \emph{spread}, or the distance between the highest bid and lowest ask price; the available \emph{liquidity}, or the total volume of shares on offer; and the \emph{distribution} of that liquidity, in particular whether it is concentrated near or far from the spread, and whether it is significantly greater on one side of the book than the other.

Our simulated limit order book follows an order matching process similar to the Nasdaq exchange in the United States.  That is, an arriving order to buy will transact with the lowest priced ask order already in the limit order book.  If the arriving order is to sell, it will transact with the highest priced bid order instead.  If there are multiple orders in the limit order book at the same price, the oldest order will be transacted.  All transactions happen at the limit price of the order already in the limit order book, not the arriving order.

\subsection{Background Traders}

Our simulated market is not a simple backtest constructed from static historical data, but rather a dynamic and interactive one in which many agents participate in pursuit of profit, each able to directly impact market pricing and other agents through its actions.  To this end we employ a population of stylized strategy agents divided into several families inspired by the work of Lawrence Harris \cite{harris1998optimal}: value-motivated traders, informed traders, and liquidity traders.  The first two represent our ``background'' trading population, which we do not alter or manipulate, and the third represents our experimental population.

Both of our representative background agent strategies obtain noisy observations of an exogenous price-time series, sometimes called the fundamental series, that represents the ``true value'' of a stock independent of current market price fluctuations.  These observations over time influence an internal value belief that differs per agent according to the update of a Bayesian process as constructed by Wah et al \cite{wah2017welfare}.  In summary, assume a background agent, arriving at the market via a Poisson process, wakes at time $t$ and receives observation $o_t$.  It updates an estimate of the current fundamental value $\tilde{r}_t$ and that estimate's variance $\tilde{\sigma}^2_t$:
\begin{equation}
  \tilde{r}_{t'} \leftarrow (1-(1-\kappa)^\delta)\bar{r} + (1-\kappa)^\delta \tilde{r}_{t'}
\end{equation}
\begin{equation}
  \tilde{\sigma}^2_{t'} \leftarrow (1-\kappa)^{2\delta} \tilde{\sigma}^2_{t'} + \frac{1-(1-\kappa)^{2\delta}}{1-(1-\kappa)^2} \sigma^2_s
\end{equation}
where $t'$ is agent's previous wake time, $\delta=t-t'$, $\kappa$ is the fundamental mean reversion parameter, and $\sigma^2_s$ is the shock variance of the fundamental process.  Having now accounted for intervening time, the agent applies its new observation $o_t$ to obtain an estimate of the current fundamental value and that estimate's variance:
\begin{equation}
  \tilde{r}_t=\frac{\sigma^2_n}{\sigma^2_n+\tilde{\sigma}^2_{t'}}\tilde{r}_{t'} + \frac{\tilde{\sigma}^2_{t'}}{\sigma^2_n+\tilde{\sigma}^2_{t'}}o_t
\end{equation}
\begin{equation}
  \tilde{\sigma}^2_t=\frac{\sigma^2_n\tilde{\sigma}^2_{t'}}{\sigma^2_n+\tilde{\sigma}^2_{t'}}
\end{equation}
where $\sigma^2_n$ is the agent's observation noise.  With updated estimates $\tilde{r}_t$ and $\tilde{\sigma}^2_t$, the agent can compute $\hat{r}_t$, the final fundamental value $r_T$ as estimated at current time $t$:
\begin{equation}
  \hat{r}_{t} \leftarrow (1-(1-\kappa)^{T-t})\bar{r} + (1-\kappa)^{T-t} \tilde{r}_{t}
\end{equation}
where $\bar{r}$ is the fundamental mean, and random perturbations are assumed to take on a mean value of zero.  This estimate of the final fundamental value represents the agent's belief about what the stock price \emph{should} be at the close of the trading day.  It uses this value to inform its decisions concerning limit price, trade direction, aggressiveness of trading posture, and so on.

Value-motivated traders tend to place limit orders away from the spread, intending to transact only if prices reach a level consistent with their private value beliefs plus a required level of surplus.  We represent this style of trading with a common variation of the Zero Inteligence (ZI) trader as described by Wah et al \cite{wah2017welfare}, which estimates the final fundamental value as explained above.  Each ZI trader is constructed with a random vector of incremental private values placed on the acquisition or release of one additional unit of stock, given the agent's current holding, which is applied as an offset to the estimated final fundamental value.  If $q_{max}$ is the holding limit, then the preferences for trading agent $i$ are the elements $\theta^q_i$ in:
\begin{equation}
  \Theta_i=(\theta^{-q_{max}+1}_i,\dots,\theta^0_i,\theta^1_i,\dots,\theta^{q_{max}}_i)    
\end{equation}
where $q$ is the quantity of stock currently held.  $\Theta_i$ is drawn randomly from $\mathcal{N}(0,\sigma^2_{PV})$, where $\sigma^2_{PV}$ is a selected experimental parameter.  The values are sorted in descending order ensuring diminishing returns on private value offsets.  The ZI trader places limit orders in a random direction (buy/sell) but selects limit prices such that transacted orders will always produce an expected surplus to the agent.

Informed traders are represented by a class of agents taking on the role of an arbitrageur.  These agents are broadly similar to the ZI agents, as they also make noisy observations of a fundamental value and construct a belief about the ``true'' worth of a stock.  The informed traders, however, always place a directional order that will profit from a reversion of the order flow to the fundamental.  Based on market conditions, these agents may place orders in one of two postures: \emph{aggressively} with market orders or limit orders that cross over the spread, or \emph{passively} with limit orders that do not cross the spread.  The supply and demand information they inject to the order flow should be predictive of short-term price moves due to their approximately correct exogenous observations.

\subsection{Experimental Traders}

Liquidity traders have no opinion about the ``correct'' value of a stock.  They participate in anticipation of profit by observing the order flows in the market for clues that suggest short-term price movement arising from the market microstructure itself.  We inject a population of liquidity traders as our experimental agents of change, observing their impact on the market, and altering their communication latency to the exchange to understand the effect this has on the profitability of their strategy.  In keeping with the spirit of liquidity trading, these agents are unable to observe the exogenous price-time series (fundamental) used by the background traders.  Our specific choice of liquidity trader follows an order book imbalance (OBI) strategy which continuously tracks what proportion of total liquidity near the spread is on the buy side of the limit order book:
\begin{equation}
  I = \frac{\sum_{b\in B}{V_b}}{\sum_{o\in O}{V_o}}
\end{equation}
where $B$ is the set of visible bid orders, $O$ is the set of all visible orders, and $V$ represents the share volume of a particular order.  For example, when the indicator $I=0.5$, liquidity is equally distributed between the two sides of the book, and when $I=0.6$ there is 50\% more bid than ask liquidity.

The agent enters a directional trade when $I>0.5+H$ or $I<0.5-H$, where $H$ is a configurable entry threshold, and exits the directional trade based on a trailing stop at configurable distance $D$ applied to the same indicator (not the midpoint stock price).  The order book depth (level) $L$ at which to consider liquidity provision to be a positively-correlated signal is also a configurable parameter.

\section{Experiments and Results}

We conducted two primary experiments for analysis in this work.  Both experiments utilize a simulated market containing a single exchange with a limit order book mechanism accepting offers to transact shares of a single security.  The experimental market is populated with 1,000 background traders, split evenly between the value-motivated and informed trader types described in Approach.

Every trading agent is configured with a minimum communication latency to the exchange.  When a message is scheduled for transmission, the \emph{time sent} is the agent's current time plus a delay that accounts for the computational complexity of the agent's strategy.  The \emph{time received} will be scheduled for the \emph{time sent} plus the minimum communication latency plus a random factor drawn from a realistic network ``jitter'' model.

For those agents requiring an exogenous price-time series, we employed a sparse mean-reverting fundamental as described in Byrd \cite{byrd2019explaining}.  This mathematical series is a continuous form of the discrete mean reverting process described in Wah et al \cite{wah2017welfare} with the addition of a second variance process which is applied at lower frequency but greater magnitude to represent infrequent ``news shocks'' that can change a trader's belief about the proper valuation of the stock.

\begin{figure}[h]
    \includegraphics[width=3in]{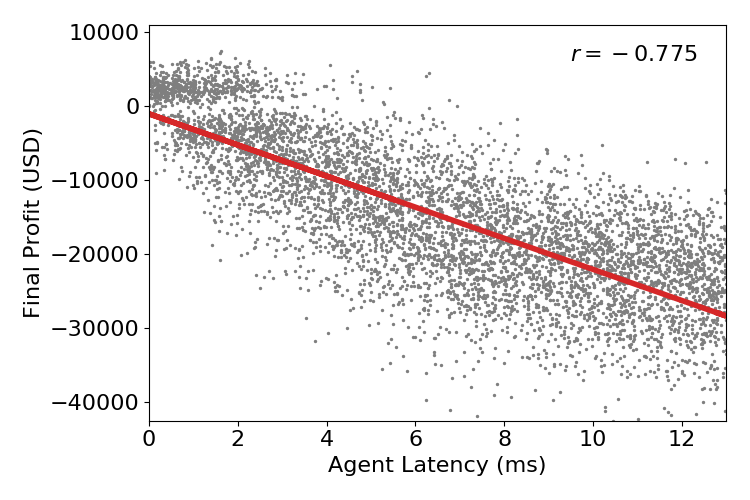}
    \caption{Profit vs absolute latency of liquidity (OBI) traders.}
    \label{fig:naive_latency}
\end{figure}

We conducted a preliminary experiment in which we added ten OBI liquidity traders, as previously described, to the background population with random latency and no particular control, and simulated hundreds of different market days.  Figure \ref{fig:naive_latency} compares the absolute latency of OBI traders to their profit at market close, with each grey data point representing one trader's result for one full market day.  We find a strong inverse Pearson correlation of $r=-0.775$.  This matches our intuition that lower latency should lead to higher profit for this strategy.  However, there is an interesting gap in the plot at low latency values, with a cluster of positive profit agents and a cluster of negative profit agents, and virtually no agents with profit close to zero.  A desire to understand this gap motivated the design of our two subsequent experiments.

\subsection{Experiment 1: Absolute Latency with Control}

In the first experiment, the described background agents were geographically situated around the United States relative to the location of the New York Stock Exchange.  The minimum communication latency of each individual background agent was drawn from a uniform distribution of 21$\mu s$ (roughly the other side of Manhattan) to 13$ms$ (around Seattle, Washington).

To the background population, we added ten order book imbalance (OBI) liquidity traders following the previously described indicator strategy.  One liquidity trader, serving as a control, was always placed at the minimum latency permitted to background agents and eight were randomly distributed in the same range as the background agents.  The final liquidity trader is considered the experimental agent for Experiment 1, for which the geographic location (latency) is systematically varied to measure its impact on the returns of all liquidity traders.  All OBI liquidity traders used an entry threshold of $H=0.17$, a trailing stop distance of $D=0.085$, and a book depth significance parameter of $L=10$.

The experiment was carried out across 600 market simulations, with each representing a full market day from 9:30 AM to 4:00 PM at nanosecond resolution.  The experimental liquidity trader was tested for a full market day at each of thirty different levels of latency, ranging from exchange colocation (approx. 333$ns$) to Seattle, Washington (approx. 13$ms$).  Each of twenty random market days (i.e. different exogenous price-time fundamental series) were repeated with the experimental agent varied among the thirty positions to minimize unintended perturbations in the market process.  We thus obtained 6,000 full-day observations of the latency and profitability relationship of an OBI liquidity trader.

\begin{figure}[h]
    \includegraphics[width=3in]{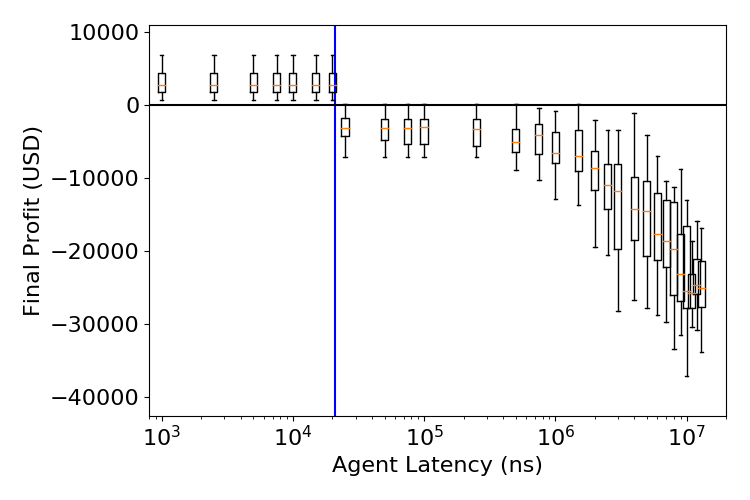}
    \caption{Profit of experimental liquidity trader for thirty levels of absolute latency (log scale) across twenty market days.  Blue line represents fixed-latency control liquidity trader.}
    \label{fig:vary_latency}
\end{figure}

\subsubsection*{Results} For this experiment, we logged the absolute latency of each liquidity trader along with its marked-to-market profit at the end of each trading day. 

Figure \ref{fig:vary_latency} is a box-plot representation of the relationship between log-scale latency in nanoseconds and final profit for the single experimental liquidity trader at each tested location.  The mean profit per tested latency is marked with a tick, the box extents mark one standard deviation, and the whiskers mark two standard deviations.  Recall that a control liquidity trader (blue line) is always placed at the minimum permitted latency outside of colocation in an effort to explain the zero profit gap in the preliminary experiment.  We can see that the experimental agent is consistently profitable when it is co-located, and performs quite poorly when far from the exchange, consistent with the earlier result.  As hoped, the combination of a control agent and systematic latency variation did add interpretive value beyond the preliminary experiment.

Let $X$ be the experimental liquidity trader with latency $L_X$ and profit $P_X$, and $C$ be the control liquidity trader with fixed latency fixed latency $L_C$ and profit $P_C$.  We now note two new observations.  First, $\forall L_X < L_C: P_X \perp L_X$.  That is, once the absolute latency of the experimental trader is lower than that of the control trader, its latency does not affect its profit.  Second, as soon as $L_X > L_C$ there is an immediate transfer from $P_X$ to $P_C$.\\

\subsubsection*{Discussion} These results together suggest that latency \emph{rank} among the competing liquidity agents is more significant to profit outcomes than absolute latency values.  This could explain the gap in the preliminary plot: Depending on whether some other agent was even closer to the exchange, a given low latency trader might see very different results.  However, we cannot support this claim based solely on Experiment 1, because the location of the \emph{other} liquidity traders (non-experimental, non-control) on each market day is not considered.

\subsection{Experiment 2: Latency Rank}

In the second experiment, ten liquidity traders were again added to the background agent population.  This time, all agents were randomly situated with a latency ranging from exchange colocation (333$ns$) to Seattle, Washington (13$ms$), with the intent to examine latency rank independent of absolute latency.  Without the need to resimulate the same market day for control, this experiment was conducted across 600 different market days, producing again 6,000 observations of OBI liquidity trader latency and profit.  The experiment was otherwise similar to Experiment 1.

\begin{figure}[h]
    \includegraphics[width=3in]{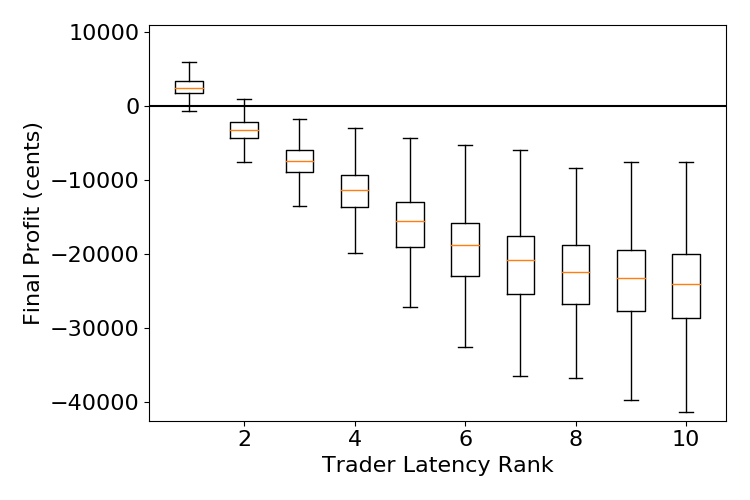}
    \caption{Profit of multiple trials with liquidity traders distributed randomly, identified by latency rank order within each trial.  Lower rank traders are closer to the exchange.}
    \label{fig:latency_rank_whisker}
\end{figure}

\subsubsection*{Results} For Experiment 2, the latency \emph{rank} of each liquidity trader was logged along with the marked-to-market profit at the end of each trading day.  For example, the liquidity trader closest to the exchange on a given day would be latency rank 1, the second closest rank 2, and so on.  Absolute latency was also logged for visualization purposes.  In Figure \ref{fig:latency_rank_whisker}, we present a box-and-whisker plot of aggregated profit by latency rank among the OBI liquidity agents.

Over the course of a simulated trading day, without consideration of absolute latency, the liquidity trader closest to the exchange receives mean and std marked-to-market profit of \$2,681.04 and \$1,389.37.  The second closest liquidity trader receives mean and std profit of \$-3,297.30 and \$1,661.30, and the liquidity trader furthest from the exchange received mean and std profit of \$-24,473.64 and \$-6,449.97.  This supports the notion that latency rank is the key factor driving the allocation of profit within this trading strategy population.

However, there is a potential confounding factor.  It \emph{could} be the case that liquidity trader latency distributions were such that the lowest ranked agent was also always close to the exchange in absolute terms.  A different look at the data will help to clarify this.

\begin{figure}[h]
    \includegraphics[width=3in]{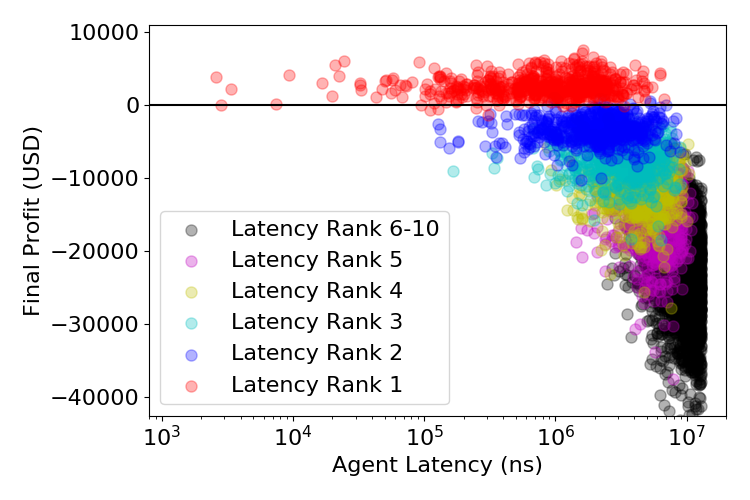}
    \caption{Profit of multiple trials with liquidity traders distributed randomly showing ranked and absolute latency.}
    \label{fig:latency_rank_raw}
\end{figure}

In Figure \ref{fig:latency_rank_raw}, we plot end of day profit against log-scale nanosecond latency, with the addition of a zero profit line.  Each point represents one liquidity trader's result for one simulated market day, and each point is color-coded according to that trader's latency rank for that simulated market day, with red being rank 1, blue being rank 2, and so on.  Ranks 6-10 are aggregated into a single black grouping.  Because this plot shows both absolute latency and latency rank together, we can see that the nearest liquidity trader (red) on a given market day does well even when situated very far from the exchange, and liquidity agents in ranks 2 and 3 perform poorly even when relatively near the exchange.

\subsubsection*{Discussion} Taking Figures \ref{fig:latency_rank_whisker} and \ref{fig:latency_rank_raw} together, we can clearly state that: The liquidity trader closest to the exchange rarely loses money, and the second closest liquidity trader rarely makes money, regardless of their absolute distance from the exchange.

Our initial concern regarding the preliminary experiment's zero-profit gap proved correct.  While there was a strong inverse correlation between absolute latency and profit, this was not the primary factor influencing strategy returns.  Profitability was rather determined by the ordinal latency \emph{rank} among agents pursuing a similar low-latency strategy.

\section*{Conclusion}

Using a realistic agent based market simulation, we explored the effect of latency on the profitability of a strategy that depends on perishable information.  We constructed an environment consisting of thousands of background trading agents into which we could inject low latency liquidity traders acting on an order book imbalance indicator.

Through a preliminary experiment, we observed a strong inverse correlation between the absolute latency and profitability of liquidity traders.  We systematically investigated this through two controlled experiments and found that the result of the preliminary experiment was not sufficiently explanatory.  While the correlation is real, the actual determining factor in profitability was the ordinal rank of latency among the population of liquidity traders.

Regardless of absolute distance, we found that the closest liquidity trader was substantially more profitable than agents even slightly further away, and was the only consistently net profitable liquidity trader.  These observed winner-take-all outcomes among traders pursuing the same time-sensitive strategy would seem to justify the apparent ``arms race'' to achieve the minimum possible latency to an exchange when engaging in market microstructure trading.

\bibliographystyle{ACM-Reference-Format}
\bibliography{biblio}

\end{document}